\documentclass[sigconf,screen,nonacm]{acmart}
\usepackage{enumitem}

\setcopyright{none}
\acmDOI{}
\acmConference[UMAP '26]{ACM UMAP 2026: 34th ACM Conference on User Modeling, Adaptation and Personalization}{June 8--11, 2026}{Gothenburg, Sweden}
\acmBooktitle{ACM UMAP 2026: 34th ACM Conference on User Modeling, Adaptation and Personalization, June 8--11, 2026}
\acmISBN{}

\begin{document}

\title[What Do Humanities Scholars Need? A User Model for Recommendation in Digital Archives]{What Do Humanities Scholars Need? \\
 A User Model for Recommendation in Digital Archives}

\author{Florian Atzenhofer-Baumgartner}
\email{florian.atzenhofer-baumgartner@student.tugraz.at}
\orcid{0000-0001-8157-8629}
\affiliation{%
  \institution{Graz University of Technology}
  \city{Graz}
  \country{Austria}\\
  \institution{University of Graz}
  \city{Graz}
  \country{Austria}
}

\author{Dominik Kowald}
\email{dkowald@know-center.at}
\orcid{0000-0003-3230-6234}
\affiliation{%
  \institution{Know Center Research GmbH}
  \city{Graz}
  \country{Austria}\\
  \institution{University of Graz}
  \city{Graz}
  \country{Austria}
}

\begin{abstract}
User models for recommender systems (RecSys) typically assume stable preferences, similarity-based relevance, and session-bounded interactions---assumptions derived from high-volume consumer contexts. This paper investigates these assumptions for humanities scholars working with digital archives. Following a human-centered design approach, we conducted focus groups and analyzed interview data from 18 researchers. Our analysis identifies four dimensions where scholarly information-seeking diverges from common RecSys user modeling: 
(1)~\emph{context volatility}---preferences shift with research tasks and domain expertise;
(2)~\emph{epistemic trust}---relevance depends on verifiable provenance;
(3)~\emph{contrastive seeking}---researchers seek items that challenge their current direction;
and (4)~\emph{strand continuity}---research spans long-term threads rather than discrete sessions. 
We discuss implications for user modeling and outline how these dimensions relate to collaborative filtering, content-based, and session-based recommendation. We propose these dimensions as a diagnostic framework applicable beyond archives to similar application domains where typical user modeling assumptions may not hold.
\end{abstract}



\maketitle

\begingroup
\renewcommand\thefootnote{}\footnotetext{%
  \hspace{-1.5em}\raisebox{5pt}{%
    \begin{minipage}[t]{\columnwidth}
      \footnotesize
      © Florian Atzenhofer-Baumgartner and Dominik Kowald, 2026. This is the author's version of the work entitled ``What Do Humanities Scholars Need? A User Model for Recommendation in Digital Archives''. It is posted here for your personal use, not for redistribution. The definitive version of record was accepted for publication in the \textit{34th ACM International Conference on User Modeling, Adaptation and Personalization (UMAP 2026)}. DOI: \url{https://doi.org/10.1145/3774935.3806171}.
    \end{minipage}%
  }%
}
\endgroup

\section{Introduction}

User modeling is foundational to adaptive systems, yet the implicit assumptions embedded in user models are typically derived from domains where engagement drives design~\cite{purificatoUserModelingUser2024, heSurveyUserBehavior2023}.
While scholarly information-seeking in digital environments has received attention~\cite{sinnHistoriansUseDigital2014a}, digital archives holding \emph{primary sources}---historical documents, charters, photographs, administrative records---remain largely unexplored from a user modeling perspective.

This gap matters for two reasons.
First, scholars working with primary sources exhibit different information-seeking behavior than users of consumer platforms~\cite{chassanoffHistoriansExperiencesUsing2018, traceInformationManagementHumanities2017}: they seek \emph{evidence} to construct arguments, their practice is comparative and critical, and their work spans long-term research projects rather than discrete sessions~\cite{owensDigitalSourcesDigital2021a}.
Second, the functional purpose of archival platforms determines how user-item interaction occurs: these systems mediate access to shared (cultural) heritage, and design choices can shape what histories can be written~\cite{zhaoValueCocreationCultural2024}.
Recent critiques have called for ``recommending with, not for'' users~\cite{ekstrandRecommendingNotCoDesigning2025c} and questioned whether standard evaluation paradigms capture domain-specific utility~\cite{saidWereStillDoing2025b}.
These calls resonate where value emerges through extended scholarly engagement rather than immediate interaction.

In this paper, we investigate common user modeling assumptions for humanities scholars.
We analyze focus group data from 18 researchers and make three contributions:
(i) we identify four dimensions along which scholarly information-seeking diverges from standard recommender system (RecSys) assumptions: \emph{context volatility}, \emph{epistemic trust}, \emph{contrastive seeking}, and \emph{strand continuity}; 
(ii) we provide empirical grounding for critiques of standard user modeling assumptions through \textit{qualitative data from focus groups with domain experts}. Building on prior stakeholder research in this archive ecosystem~\cite{atzenhofer-baumgartnerValueIdentificationMultistakeholder2024b, atzenhofer-baumgartnerChallengesImplementingRecommender2024a}, we extend stakeholder analysis toward a domain-specific user model; 
(iii) we propose \textit{design implications} and a \textit{diagnostic framework} for identifying when standard user modeling assumptions are likely to fail in specialized domains.

\section{Related Work and Background}

Purificato et al.~\cite{purificatoUserModelingUser2024} provide definitions clarifying \emph{user modeling} as the process of acquiring and representing user characteristics, noting paradigm shifts toward implicit data and multi-behavior modeling.
However, these frameworks assume high-volume consumer contexts where preference stability is a core assumption.
This assumption raises questions about external validity: Wardatzky et al.~\cite{wardatzkyWhomExplanationsServe2025c} find that user studies in explainable Recommender Systems (RecSys) predominantly cover participants who may not represent actual system users in the evaluation domain.
These gaps are particularly visible in specialized domains, where user models must account for varying expertise levels~\cite{kostricShouldWeTailor2025} and, potentially, the growth of proficiency over time.

Information science has long studied scholarly information-seeking, revealing patterns distinct from general web search~\cite{traceInformationManagementHumanities2017, chassanoffHistoriansExperiencesUsing2018}.
A key distinction is organizational: digital archives differ from libraries in their focus on primary sources and provenance-based or tectonic arrangement~\cite{owensDigitalSourcesDigital2021a}.
This provenance orientation shapes search behavior; researchers engage in ``berrypicking''~\cite{Bates_1989}---iteratively refining queries as understanding develops~\cite{Savolainen_2018}.
Beyond search patterns, digital archives present infrastructural challenges: metadata quality varies, and users require contextual understanding of record groups and their relationships~\cite{latePerfectWorldExploring2023, matusiakEvaluatingDigitalCommunity2022}.
In light of cultural heritage, these temporal and expertise-dependent patterns contrast with RecSys applications in museums, which commonly address single-visit optimization~\cite{Casillo_Colace_Conte_Lombardi_Santaniello_Valentino_2023}; archival research instead involves extended timelines and non-linear progressions~\cite{Li_Zhang_Wang_2024}.

Recent work has challenged the 'designer-centric' paradigm in RecSys.
Ekstrand et al.~\cite{ekstrandRecommendingNotCoDesigning2025c} argue that systems should be designed \emph{by} and \emph{with} users, not merely \emph{for} them. 
Such concerns become especially acute when considering concepts like serendipity: Binst et al.~\cite{binstWhatSerendipityInterview2025c} conceptualize experienced serendipity as having fortuitous, refreshing, and enriching components.
Despite increasing calls for participatory approaches, empirical grounding through qualitative research with domain experts remains rare.
Kostric et al.~\cite{kostricShouldWeTailor2025} provide one such grounding, showing how user expertise affects preference elicitation in scientific literature recommendation.

Our work extends this line by grounding user model dimensions in qualitative data from focus groups with 18 humanities researchers, building on our prior multistakeholder research in this archive ecosystem~\cite{atzenhofer-baumgartnerValueIdentificationMultistakeholder2024b, atzenhofer-baumgartnerChallengesImplementingRecommender2024a} and analyzing their statements specifically for user modeling assumptions.

\section{Methods}

\noindent \textbf{Focus Group Design.} 
We conducted five 60-minute focus groups with stakeholders involved in a large digital archive ecosystem hosting primary historical sources---specifically medieval charters, each comprising a digitized document image and structured metadata (text, abstract, date etc.).
Following calls for multistakeholder evaluation in RecSys~\cite{Burke_Adomavicius_Bogers_Noia_Kowald_Neidhardt_Özgöbek_Pera_Tintarev_Ziegler_2025, burkeDagstuhlSeminarEvaluation2024}, the broader study involved 25 domain experts across five stakeholder groups: upstream (archivists, curators), provider (aggregators, digitization services), system (developers, managers), consumer (researchers, educators), and downstream (publishers, platforms).
From these 25 participants, we identified 18 who were actively engaged in research based on (a) current involvement in scholarly projects using digital archives and (b) regular use of archival platforms for research purposes. This subset included participants from all five stakeholder groups (5 upstream, 2 provider, 2 system, 5 consumer, and 4 downstream), ensuring perspectives from both content creators and content users regarding research practice.
For this paper, we analyze their statements specifically regarding information-seeking behavior.

Focus groups followed a semi-structured protocol with three thematic topics: (1)~visibility and representation, (2)~adaptation and access, and (3)~transparency and trust.
Each topic included scenario-based mockups illustrating different recommendation approaches (e.g., popularity-based vs.\ diversity-based vs.\ personalized) to ground discussion in concrete design choices.
Provocative statements (e.g., ``I don't need to understand why something is recommended as long as it's relevant to my research'') elicited varied value positions.
Participants received background materials on recommender systems and user modeling prior to the sessions; prior experience with RecSys varied across participants.
Sessions were recorded and transcribed.\footnote{All participants provided written informed consent following institutional ethical guidelines. Interview protocols and materials available at: \url{https://github.com/atzenhofer/user-modeling-archives-recsys}}

\vspace{1mm} \noindent \textbf{Analysis.} 
We performed thematic analysis on transcripts using an abductive approach, combining deductive coding from RecSys literature with inductive analysis of emerging patterns.
Our coding focused specifically on statements revealing: (1)~expectations and assumptions about recommendation behavior; (2)~references to other recommendation systems; and (3)~descriptions of information-seeking needs that conflicted with standard RecSys assumptions (e.g., wanting dissimilar items, distrusting popularity signals).
Through iterative coding, we identified four recurring themes that structure scholarly information-seeking in ways that diverge from commercial user modeling.

\section{Results and Findings}

Our analysis identifies four dimensions where scholarly information-seeking behavior diverges from typical user modeling assumptions.
Conceptually, these dimensions manifest at different scopes of the user-system interaction: \emph{context volatility} (depending on the user's internal state, task, and expertise); \emph{epistemic trust} (understood as acceptance prerequisite and with provenance verification); \emph{contrastive seeking} (referring to relational intent and momentary deflection); and \emph{strand continuity} (governed by a temporal lifecycle, as part of a (long-term) research thread).
Together, they form a framework for modeling specialized information needs.

\subsection{Context Volatility}

\noindent \textbf{Standard Assumption.} User preferences are stable and can be modeled from historical behavior.

\vspace{1mm} \noindent \textbf{Observed Behavior.} Preferences shift rapidly based on current research tasks and domain expertise.
Researchers switch between projects, topics, and personas (e.g., ``teacher mode'' vs.\ ``researcher mode'').
Furthermore, user models must account for the \emph{growth of expertise} over time. As scholars become increasingly proficient with archival material and platform navigation, their information-seeking patterns evolve from broad exploration to targeted verification. Expertise itself is context-dependent:

\begin{quote}
\emph{``You might be easily considered an expert in your own language... but then there might be a third, fourth and fifth language where you can maybe guess what is shown but are not really able to understand... you might easily become a non-expert.''}~(I11)
\end{quote}

\noindent This iterative, context-dependent nature also manifests in search behavior:

\begin{quote}
\emph{``If I go [to the archive] and ask for Charlemagne, and that's not really what I want but it's something in the near vicinity of my interest, in two or three steps I get what I really want.''}~(I7)
\end{quote}

A user profile built from one research phase may actively hinder the next.
Participants explicitly requested controllability to manage these shifts:

\begin{quote}
\emph{``Do I have, as a user, the possibility to switch---show me expert recommendations, or show me narrative recommendations?''}~(I2)
\end{quote}

This challenges collaborative filtering (CF) approaches that aggregate long-term preferences and CB approaches that build stable feature profiles.
Session-based models capture some context-dependence but miss the extremely long \emph{project} structure organizing scholarly work.

\subsection{Epistemic Trust}

\noindent \textbf{Standard Assumption.} Trust derives from social proof or system authority. Explainability provides transparency.

\vspace{1mm} \noindent \textbf{Observed Behavior.} Trust requires verifiable provenance.
Scholars need to understand \emph{why} an item was recommended to verify the reasoning themselves.
Explainability constitutes not merely transparency---here it is a prerequisite for relevance.
Participants stress that transparency is a professional obligation:

\begin{quote}
\emph{``What's the use of a recommendation when I don't understand it? So maybe I can think that the whole system is making fun of me. And as an archivist, trust and transparency is a key value for us.''}~(I15)
\end{quote}

\noindent For some participants, provenance-based presentation inherently satisfies this requirement:

\begin{quote}
\emph{``The presentation according to provenance and authority is the most important, most reliable, and it explains also by itself why documents are recommended.''}~(I17)
\end{quote}

\noindent Others emphasize that the ``why'' must be independently verifiable:

\begin{quote}
\emph{``To me the essential thing, the priority, is the `why'... if I can see the reason is relevant then okay... and then go and check the authority.''}~(I8)
\end{quote}

\noindent Failed epistemic trust leads to disengagement: \emph{``I frankly ignore recommender systems because it's very rare that the reason it was recommended is a reason that's relevant to me''}~(I8).
Popularity-based signals face particular resistance, with participants comparing them to the \emph{``TikTok sensation where one video can suddenly have millions... while others are completely buried''}~(I9).

This skepticism extends to AI-generated content and metadata---a concern echoed in archival practice~\cite{yacoWhatCanAI2025, cushingHowWeBalance2023a}, where participants worried that AI hallucinations could be ``devastating'' for scholarly platforms.

\subsection{Contrastive Seeking}

\noindent \textbf{Standard Assumption.} Users want items similar to those previously engaged with (similarity, homophily).

\vspace{1mm} \noindent \textbf{Observed Behavior.} Scholars deliberately seek items that contrast with or challenge their current direction.
Perceived controllability is crucial here: users want the power to influence how they experience recommendations, potentially switching between ``challenge me'' and ``confirm me'' modes depending on their current research phase.

\begin{quote}
\emph{``Could you not have a system which would give you synonyms and antonyms? In other words, that you would get both those recommendations confirming your research direction and those that challenge you.''}~(I3)
\end{quote}

This frames contrastive seeking as \emph{epistemic hygiene}---a disciplinary need to avoid confirmation bias, not simple preference.
Participants emphasized not wanting more results, but \emph{better} ones---defined by argumentative value:

\begin{quote}
\emph{``I don't want to find more, but maybe find better. So, find the charters or the things that I really need.''}~(I4)
\end{quote}

This comparative orientation is methodologically grounded.
Scholars described their practice as inherently contrastive:

\begin{quote}
\emph{``The most prominent method in diplomatics is comparing one charter with another charter. So if you have some recommended charters, you want to compare them, why are they recommended and what have they in common?''}~(I5)
\end{quote}

This inverts the similarity optimization at the core of CF and CB.
Serendipity in this context is an \emph{enrichment that advances the research argument}---aligning with but extending Binst et al.'s framework~\cite{binstWhatSerendipityInterview2025c}.
As one participant noted: \emph{``These sort of accidental findings are often the most rewarding ones, so, the idea of serendipity.''}~(I16)

\subsection{Strand Continuity}

\noindent \textbf{Standard Assumption.} The session is the natural unit of analysis.

\vspace{1mm} \noindent \textbf{Observed Behavior.} Work spans long-term \emph{research strands}---projects that persist for months with intermittent activity.
This dimension acts as a bridge for controllability: overarching themes can be handled explicitly through project-based modeling or implicitly through strand-aware continuity.

\begin{quote}
\emph{``Very often, these recommended things do not focus on what I'm really searching in the moment, but add something which opens new paths for my research. So it must not be so very concentrated on what I'm doing in the very, very moment.''}~(I5)
\end{quote}

This reflects the ``berrypicking'' pattern~\cite{Bates_1989}, possibly spanning months.
Participants explicitly requested session persistence and context-switching capability:

\begin{quote}
\emph{``Maybe you can have an about section where you actually explain behind the scenes how this tool is working... basically you are now interested in a different topic you don't want to have the same recommendations as before, start from anew.''}~(I9)
\end{quote}

\begin{quote}
\emph{``Would the system maintain where you were, so that you can come back the next day or a week later, and it would have your previous discussion at hand?''}~(I3)
\end{quote}

This persistence expectation connects to long-term trust: \emph{``If I get the impression that the system is learning with my input, I would make the effort to put in my opinion. If it's for nothing, I would stop''}~(I1).
Recommendations should be ``strand-aware,'' recognizing that a user's current session may relate to one of several ongoing research threads.

\section{Discussion}

\noindent \textbf{Implications for User Modeling.} 
These dimensions suggest that deploying standard approaches in scholarly archives risks fundamental misalignment.
Notably, dimensions were emphasized differently across stakeholder roles. \emph{Epistemic trust} was particularly salient among participants from upstream roles (archivists, curators) who framed transparency as a professional obligation, while \emph{contrastive seeking} emerged most strongly from consumer-researchers whose disciplinary methods require comparison. This suggests user models may benefit from role-sensitive calibration even within the scholarly domain.
Prior multistakeholder work on this archive~\cite{atzenhofer-baumgartnerMultistakeholderApproachValuedriven2025b} organized stakeholder concerns around research funnel stages (discovery, interaction, integration, impact); our dimensions complement this framing, with each dimension potentially connecting to multiple stages. We elaborate on implications for three core RecSys paradigms:

\textit{Collaborative filtering} faces particular challenges.
CF relies on the assumption that similar users share preferences. In our case, the \emph{contrastive seeking} dimension shows scholars actively want items that \emph{diverge} from their current direction.
User behavior modeling surveys~\cite{heSurveyUserBehavior2023} note the field's focus on learning interest representations from interaction histories, yet such approaches may not transfer to contexts where users actively seek unfamiliar material.
Moreover, popularity-based signals, that are fundamental to many CF approaches, are viewed critically by our participants; this aligns with broader concerns that popularity bias can limit recommendation value in contexts where discovery and novelty matter~\cite{klimashevskaiaSurveyPopularityBias2024}.
The ``users like you'' framing provides no epistemic value when researchers require verifiable provenance.
As one participant noted: \emph{``Any level of transparency will build trust... it has to be different than Amazon, which [usually] doesn't tell you why''}~(I7).

\textit{Content-based approaches} fare better on explainability since recommendations can be justified through item features.
However, CB naturally optimizes for similarity---items matching past preferences---when scholars may need contrast.
Recent work on multi-interest user modeling~\cite{lianMultiInterestAwareUserModeling2021} acknowledges that ``traditional models tend to encode a user's behaviors into a single embedding vector, which do not have enough capacity to effectively capture diverse interests.''
Our \emph{context volatility} dimension reinforces this concern: scholars might not have (single) stable profiles but volatile research contexts.

\textit{Session-based models}~\cite{wangSequentialSessionbasedRecommendations2022} capture context-dependence within bounded interactions but assume session boundaries are meaningful.
Our \emph{strand continuity} dimension points out that scholarly work organizes around long-term research strands, not sessions.
Neither session-based models (which reset context too frequently) nor full user models (aiming to aggregate everything) capture this structure.

\vspace{1mm} \noindent \textbf{Recommender Systems Design Implications.} 
While our findings emerge from digital archives, they align with challenges in other domains.
We observe that \emph{context volatility} may parallel music recommendation, where users exhibit distinct discovery patterns depending on their current needs---patterns that cannot be aggregated into a single stable model~\cite{moscatiFamiliarizingMusicDiscovery2025}.
\emph{Epistemic trust} may connect to news recommendation, where multi-stakeholder research reveals gaps between stated and revealed preferences, and where beyond-accuracy measures must balance editorial standards with user needs~\cite{kolbBridgingPreferencesMultiStakeholder2025}.
\emph{Contrastive seeking} relates to serendipity research~\cite{binstWhatSerendipityInterview2025c, Nalis_Sippl_Kolb_Neidhardt_2024}, though we argue that scholarly serendipity is not ``pleasant surprise'' but \emph{enrichment that advances a research argument}.
\emph{Strand continuity} may share structure with project-based work broadly, where users maintain long-term investigative threads.

We argue that a key differentiator is that scholarly work with primary sources combines all four dimensions simultaneously. Context volatility alone occurs in other domains, but here it co-occurs with methodologically driven contrastive seeking, trust that demands verifiable provenance rather than social proof, and strand continuity spanning months or years.
This suggests our dimensions may serve as a \textit{diagnostic framework}: when deploying RecSys in a new domain, practitioners should assess whether users exhibit (1)~task-dependent preference shifts, (2)~verification-based trust requirements, (3)~deliberate divergence-seeking, or (4)~'project-spanning' interaction patterns. Presence of multiple dimensions signals potential misalignment with standard approaches.

Concrete design directions include:
(1)~supporting explicit ``project'' or ``strand'' contexts that persist across sessions;
(2)~offering contrastive recommendations alongside similar ones---a ``challenge me'' mode;
(3)~providing provenance metadata as a first-class ranking signal, not optional detail;
(4)~allowing users to ``pause'' or ``reset'' recommendation contexts when switching tasks.

These align with trustworthy RecSys research~\cite{Ge_Liu_Fu_Tan_Li_Xu_Li_Xian_Zhang_2025}, which discusses controllability as a component of system trustworthiness.
Studies on user perceptions of control~\cite{Ghori_Dehpanah_Gemmell_Mobasher_2025} reveal users often perceive only an ``illusion of choice''---reinforcing our participants' requests for meaningful controllability.
Recent work on social media platforms~\cite{Li_Kuo_Sheng_Zhang_Wu_2025} identifies ``intentional implicit feedback''---deliberate user actions to shape recommendations that fall outside traditional explicit mechanisms---suggesting users develop sophisticated strategies to influence algorithmic outputs.
Our participants' explicit requests for mode-switching (``show me expert recommendations'') and context reset (``start from anew'') suggest that superficial control mechanisms are insufficient; scholars require agency over recommendation scope and direction.
Promising directions include open user models with explanations~\cite{hendrawanExplanationsOpenUser2024}, knowledge graphs exposing provenance, and multi-context models per project.

\section{Conclusion and Future Work}

Based on focus group data from 18 researchers actively engaged with digital archives, we identify four dimensions where scholarly information-seeking diverges from standard user modeling assumptions: context volatility, epistemic trust, contrastive seeking, and strand continuity.
These dimensions provide a \textit{diagnostic framework} for assessing assumption misalignment~\cite{saidWereStillDoing2025b, bauerExploringLandscapeRecommender2024b} and a foundation for designing systems that treat provenance as constitutive of relevance and model long-term research strands.

Our findings are exploratory and based on qualitative data from a specific domain; since no recommender system is currently deployed in this ecosystem, log-based behavioral validation remains future work.
Open questions include stopping conditions, negative preference expression, and social reluctance around revealing research interests~\cite{purificatoUserModelingUser2024}.
We propose to apply this framework to other specialized domains where standard assumptions about user-item interaction may similarly fail.
\begin{acks}
This research is supported by the ERC Advanced Grant project (101019327) ``From Digital to Distant Diplomatics'' and the Austrian FFG COMET program. Special thanks to the Monasterium.net and ICARus team for providing invaluable feedback and support.
During the preparation of this work, the authors used Claude 4.5 Opus to improve phrasing and flow of existing content.
\end{acks}

\bibliographystyle{ACM-Reference-Format}
\bibliography{references}

\end{document}